\documentclass[twocolumn,showpacs,preprintnumbers,amsmath,amssymb]{revtex4}

\usepackage{graphicx}
\usepackage{dcolumn}
\usepackage{bm}

\newcommand{\be}[1]{\begin{equation}\label{eq:#1}}
\newcommand{\ee}{\end{equation}}
\newcommand{\bea}{\begin{eqnarray}}
\newcommand{\eea}{\end{eqnarray}}

\newcommand{\bt}{\textbf}

\newcommand{\phd}{\phantom{\dag}}
\newcommand{\ph}{\phantom{.}}
\newcommand{\up}{^{\phd}}
\newcommand{\noi}{\noindent}

\begin{document}
\def\v#1{{\bf #1}}

%\preprint{APS/123-QED}

\title{Meissner effect without superconductivity from a chiral d-density wave}

\author{P. Kotetes}\email{pkotetes@central.ntua.gr}
\author{G. Varelogiannis}\email{varelogi@central.ntua.gr}
\affiliation{Department of Physics, National Technical University
of Athens, GR-15780 Athens, Greece}

\vskip 1cm
\begin{abstract}
We demonstrate that the formation of a chiral d-density wave
(CDDW) state generates a \textit{Topolo\-gical Meissner effect}
(TME) in the absence of any kind of superconductivity. The TME is
identical to the usual superconducting Meissner effect but it
appears only for magnetic fields perpendicular to the plane while
it is absent for in plane fields. The observed enhanced
diamagnetic signals in the non-superconducting pseudogap regime of
the cuprates may find an alternative interpretation in terms of a
TME, originating from a chiral d-density wave pseudogap.
\end{abstract}

 \maketitle

The Meissner effect is considered to be the most direct signature
of superconductivity \cite{Schrieffer}. However, the surprising
observations of such enhanced diamagnetic signals \cite{Exp Nerst
Magnet} well above the superconducting transition temperature in
the pseudogap regime of the cuprates \cite{pseudogap review},
constitute a fascinating puzzle. There are two proposals for the
nature of this regime that appear to dominate. The first,
associates the pseudogap with a $d_{x^2-y^2}$ density wave (DDW)
\cite{Chakravarty,Affleck}, also called orbital anti\-ferromagnet
\cite{DDW,Nersesyan}, which normally competes with
superconductivity. The second associates the pseudogap with
spontaneous vortex-antivortex unbinding leading to incoherent
superconductivity \cite{Emery} that should persist well above the
superconducting $T_c$. This theory is reminiscent of the well
known Kosterlitz-Thouless transition \cite{KT}.

The available ARPES \cite{ARPES} and STM \cite{STM} experiments
cannot differentiate a SC from a density wave (DW) gap, and
therefore appear somehow incapable in settling directly the issue.
On the other hand, the unusual Nernst effect and most importantly,
the enhanced {\it diamagnetic} signal that accompanies it for a
very large temperature region above the SC critical temperature
\cite{Exp Nerst Magnet}, has been considered as a major argument
in favor of the incoherent SC scenario. In fact, the enhanced
diamagnetism is viewed as a signature of the usual Meissner effect
associated solely with the SC state, and would therefore
contradict the $d_{x^2-y^2}$ density wave scenario since no
Meissner effect was expected in that case \cite{Nersesyan}.

In this letter we put forward
% a novel type of Meissner effect,
 the
Topological Meissner effect (TME), that results from a chiral
$d_{xy}+id_{x^2-y^2}$ density wave (CDDW) state. In fact, the
Nernst region of the pseudogap regime may well be associated with
a CDDW. The most intriguing property of a CDDW is that parity
(${\cal P}$) and time-reversal (${\cal T}$) violation induces
Chern-Simons terms in the effective action of the electromagnetic
field, providing the possibility of the TME and the Spontaneous
Quantum Hall effect (SQHE) earlier discussed
\cite{KVL,Yakovenko,Kerler,Goryo,Furusaki,Horovitz,Volovik}. As we
shall demonstrate, the TME is described by \textit{the same}
equation we find in the usual Meissner effect of a superconductor.
Though, its origin is radically different. In our system we
encounter the realization of Parity Anomaly
\cite{Yakovenko,parityanomaly}, with the emerging Chern-Simons
terms provi\-ding a topological mass to the electromagnetic field,
in a gauge invariant manner \cite{Deser,Fradkin}. Moreover, the
possession of chirality perpendicular to the plane, implies that
the TME is strongly anisotropic. Particularly, it takes place for
magnetic fields \textit{perpendicular} to the plane while it is
\textit{absent for in plane fields}, in accordance with the
experimental observations \cite{Exp Nerst Magnet}. Note finally
that a chiral d-density wave state has also been shown recently
\cite{Tewari} to explain the experimental results concerning the
Polar Kerr effect in YBCO \cite{Xia}.

In order to demonstrate how the TME arises, we shall consider the
following BCS hamiltonian for the CDDW \bea{\cal H}_{ {\small
CDDW}}^{\phd}=\frac{1}{2}\sum_{\bm{k}}\left(\Delta_{\bm{k}}\up
c_{\bm{k}}^{\dag}
c_{\bm{k+Q}}^{\phd}+\Delta^{*}_{\bm{k}}c_{\bm{k+Q}}^{\dag}
c_{\bm{k}}^{\phd}\right),\label{eq:Hdw}\eea

\noi which describes a $d_{xy}+id_{x^2-y^2}$ state characterized
by the wave-vector $\bm{Q}=(\pi,\pi)$, which is
\textit{commensurate} to the lattice ($\bm{k}+2\bm{Q}=\bm{k}$).
Since spin degrees of freedom do not get involved we have
considered spinless electrons, so that all our results will refer
to one spin component. Furthermore, we use
$g_{\mu\nu}\up=(1,-1,-1)$, $k^i=\bm{k}=(k_x\up,k_y\up)$,
$k^{\mu}=k=(\omega,\bm{k}$), $q^{\mu}=q=(q_0\up,\bm{q})$,
$\mu=0,1,2$, $i=1,2$, $e>0$, $\hbar=1$ and we assume that repeated
indices are summed. In the derivation of the Chern-Simons terms we
shall restrict ourselves to the zero temperature case while
necessary extensions to finite temperatures will be afterwards
performed. In addition, the summation in $\bm{k}-$space is all
over the whole $1^{st}$ Brillouin zone rather than the reduced
Brillouin zone. This implies that the operators $c_{\bm{k}}\up$
and $c_{\bm{k+Q}}\up$ do not describe independent degrees of
freedom.
\par In Eq.(\ref{eq:Hdw}) we have introduced the CDDW order
parameter $\Delta_{\bm{k}}\up=\eta\Delta\sin k_x\up\sin
k_y\up+i\Delta\left(\cos k_x\up-\cos k_y\up\right)$, where
$\Delta$ is the modulus of the $id_{x^2-y^2}$ order parameter,
$\eta$ defines the relative magnitude of the two components and
also determines the direction of the chirality of the state. The
chiral character of the state implies the existence of an
intrinsic angular momentum in $\bm{k}-$space, perpendicular to the
plane, originating from ${\cal P-T}$ violation. Specifically, the
$d_{x^2-y^2}$ component violates ${\cal T}$ as it is imaginary,
while the $d_{xy}$ component is odd under ${\cal P}$ in two
dimensions, which is defined as $(k_x,k_y)\rightarrow(k_x,-k_y)$.

In order to obtain the total electronic Hamiltonian ${\cal H}$, we
have to add the corresponding kinetic part ${\cal H}_{kin}\up$.
For the kinetic part we keep only the nearest neighbors hopping
term $\epsilon_{\bm{k}}\up=-t\left(\cos k_x\up+\cos k_y\up\right)$
satisfying the nesting condition
$\epsilon_{\bm{k+Q}}\up=-\epsilon_{\bm{k}}\up$, while we also set
the chemical potential equal to zero. Our approximation can be
justified by considering that our system is close to half-filling.
Under these conditions the excitation spectrum consists of two
bands which are fully gapped leading to the topological
quantization of the Hall conductance
\cite{KVL,Yakovenko,Kerler,Goryo,Furusaki,Horovitz,Volovik}, which
is the coefficient of the Chern-Simons terms. Omitting the next
nearest neighbors hopping term $\delta_{\bm{k}}\up=t'\cos k_x\cos
k_y$ does not alter qualitatively the occurrence of the TME.
However, its inclusion would destroy the quantization of the Hall
conductance, as in this case, the system is not fully gapped.
Similar effects would arise in the presence of disorder or by
including the $z$-axis hopping term.

Under this conditions, the total Hamiltonian of the system becomes
$ {\cal H}=\frac{1}{2}\sum_{\bm{k}}\left[\epsilon_{\bm{k}}\up
\left(c_{\bm{k}}^{\dag}c_{\bm{k}}\up-c_{\bm{k+Q}}^{\dag}c_{\bm{k+Q}}^{\phd}\right)+\right.$
$\left.\left(\Delta_{\bm{k}}\up c_{\bm{k}}^{\dag}
c_{\bm{k+Q}}^{\phd}+h.c.\right)\right]$. We obtain a compact
representation of ${\cal H}$ by introducing the spinor
$\Psi^{\dag}_{\bm{k}}=\frac{\ph 1}{\sqrt{2}}(c^{\dag}_{\bm{k}} \ph
c^{\dag}_{\bm{k+Q}})$, the isospin Pauli matrices $\bm{\tau}$ and
the vector $\bm{g}_{\bm{k}}\up\equiv({\cal
R}e\Delta_{\bm{k}}\up,-{\cal
I}m\Delta_{\bm{k}}\up,\epsilon_{\bm{k}}\up)$. This yields ${\cal
H}=\sum_{\bm{k}}\Psi^{\dag}_{\bm{k}}\ph\bm{g}_{\bm{k}}\up\cdot\bm{\tau}\ph\Psi_{\bm{k}}\up$.
The latter indicates that the ground state of the system depends
on the orientation of the $\bm{g}$ vector in isospin space. As a
result, this hamiltonian supports skyrmion solutions which imply
the presence of a Chern-Simons action (see e.g. \cite{Volovik}).

To reveal the emerging Chern-Simons terms, we have to take into
account the fluctuations of the $U(1)$ gauge field $A^{\mu}$. We
add to the Hamiltonian the term ${\cal
H}_{em}\up=\int\frac{\rm{d}^2q}{(2\pi)^2}\sum_{\bm{k}}\Psi_{\bm{k+q}}^{\dag}\Gamma^{\mu}_{\bm{k+q,k}}
A_{\mu}\up(\bm{q})\Psi_{\bm{k}}\up
-\int\frac{\rm{d}^2q}{(2\pi)^2}\sum_{\bm{k}}\Psi_{\bm{k+q}}^{\dag}\frac{e^2}{2m}A^i(-\bm{q})A_i\up(\bm{q})
\Psi_{\bm{k}}\up$, which describes the interaction of the gauge
field with the electrons. We have introduced the paramagnetic
interaction vertex $\Gamma^{\mu}_{\bm{k+q},\bm{k}}=-(e\ph,\ph
e\frac{\partial\phd}{\partial
k^i}\ph\bm{g}_{\bm{k}}\up\cdot\bm{\tau})$, where $\mu=0,1,2$ and
$i=1,2$. At one-loop level, the effective action $S_{em}\up$ is
given by the relation $S_{em}^{\phd}=\frac{1}{2}\int \frac{{\rm
d}^3q}{(2\pi)^3}\ph A^{\mu}(-q)\Pi_{\mu\nu}(q)A^{\nu}(q)$, with
the Polarization tensor $\Pi_{\mu\nu}$, defined as $
\Pi^{\mu\nu}(q)=\frac{i}{2}\int_kTr\left({\cal
G}_k\up\Gamma^{\mu}_{k,k+q}{\cal
G}_{k+q}\up\Gamma^{\nu}_{k+q,k}\right)-\frac{\ph
e^2}{m}\rho_e\up\delta_{i,j}\up$. $\rho_e\up$ is the
two-dimensional electronic density (without including spin), $Tr$
denotes trace over isospin indices, ${\cal G}_k\up$ is the CDDW
fermionic propagator and we have used the abbreviation
$\int_k=\int\frac{{\rm d}\omega}{2\pi}\sum_{\bm{k}}$. Computing
$\Pi_{\mu\nu}$ up to linear order in $q$, yields the Chern-Simons
action \bea S_{CS}\up=\int {\rm
d^3}x\ph\frac{\sigma_{xy}}{4}\varepsilon_{\mu\nu\lambda}^{\phd}A^{\mu}F^{\nu\lambda},\label{eq:Scs}\eea

\noi with
$F_{\mu\nu}=\partial_{\mu}A_{\nu}-\partial_{\nu}A_{\mu}$. The
coefficient of the Chern-Simons action is the Hall conductance
$\sigma_{xy}$. It can be shown that it is a topological invariant,
reflecting the existence of a topologically non trivial, ${\cal
P-T}$ violating ground state (see e.g. \cite{Volovik}). Using
Eq.(\ref{eq:Scs}) we obtain
\bea\sigma_{{{xy}}}^{\phd}=\frac{i}{2!}\varepsilon_{0ji}\frac{\partial\Pi_{0i}^{\phd}}{\partial
q^j}=\frac{e^2}{4\pi}\widehat{N}=\frac{e^2}{2\pi},\eea

\noi where we have introduced the winding number of the unit
vector
$\hat{\bm{g}}_{\bm{k}}\up={\bm{g}}_{\bm{k}}\up/{\mid\bm{g}_{\bm{k}}\up\mid}$,

\bea\widehat{N}=\frac{1}{4\pi}\int {\rm d}^2k\phd
\hat{\bm{g}}_{\bm{k}}\up\cdot\left(\frac{\partial\hat{\bm{g}}_{\bm{k}}\up}{\partial
k_{{x}}}\times\frac{\partial\hat{\bm{g}}_{\bm{k}}\up}{\partial
k_{{y}}}\right),\eea \noi which is equal to 2, because the order
parameter components are eigenfunctions of the angular momentum in
$\bm{k}-$space with eigenvalue $l=2$.

In the case of a perfect gap, the Hall conductance originates only
from the chirality $\widehat{N}$ of the lower energy band,
$E_{\bm{k}}^-=-|g_{\bm{k}}\up|$, which is fully occupied. In the
same time, the upper band, $E_{\bm{k}}^+=+|g_{\bm{k}}\up|$, is
totally empty while it is characterized by opposite chirality.
Apparently, if both bands were equally occupied then
$\sigma_{xy}\up$ would be equal to zero. In the general case, the
two bands, have different occupation numbers $n_-$ and $n_+$,
yielding a non-quantized Hall conductance
$\sigma_{xy}=\frac{e^2}{2\pi}(n_--n_+)$. Deviations from nesting,
disorder or a chemical potential generally lead to such an effect.
It is desirable to comprehend, even crudely, the effect of these
parameters on the Hall conductance and the TME.

For this purpose we consider that a finite chemical potential is
added to the system. We shall consider that its magnitude is of
the order of $min|g_{\bm{k}}\up|$. This minimum is realized at the
points $\bm{k}_0=(\pm\frac{\pi}{2},\pm\frac{\pi}{2})$, when
$\eta<<1$. In this case, we may linearize the spectrum about these
points so to obtain an approximate analytical solution. The two
energy bands are described by the dispersions
$E_{\bm{k}}^{\pm}=-\mu\pm\sqrt{m^2+(\bm{v}_0\cdot\delta\bm{k})^2}$,
with $m=min|g_{\bm{k}}\up|=|g_{\bm{k}_0}\up|$, $\bm{v}_0$ the
velocity at these points and $\delta\bm{k}=\bm{k}-\bm{k}_0\up$. If
$|\mu|\geq m$ and $\mu<0$, hole-pockets arise in the lower band
decreasing the full occupancy from $n_-=1$ to $n_-=1-n_{ex}$, with
$n_{ex}$ the portion of the empty states. On the other hand, if
$\mu\geq m$, electron pockets emerge in the upper band rising its
occupancy from zero. However, if we take into consideration that
the two bands have opposite chirality, it is evident that in both
cases, the effect is the same. Consequently,
$\sigma_{xy}(\mu)=\sigma_{xy}(1-n_{ex})$. The portion of the empty
states will be determined by the area of the ellipses defined by
the four hole-pockets. Straightforward calculations yield the
simple relation \bea n_{ex}=(\mu^2-m^2)/2\pi
t\Delta\,.\label{eq:nex}\eea \noi We observe that for small values
of $|\mu|$, compared to $t$ and $\Delta$, the effect of doping is
negligible.

We are now in position to obtain the equations of motion of the
gauge field which will allow us to discuss the TME in a Hall bar
geometry setup. We consider that the Hall bar has dimensions
$L_x=2l_x, L_y>>L_x$ extending from $-l_x$ to $l_x$ on the
$x$-axis. The relation $L_y>>L_x$ indicates that there is
negligible $y$-dependence of the gauge fields ($\partial_y=0$). To
describe the dynamics of the propagating gauge field we have to
add in Eq.(\ref{eq:Scs}) the three-dimensional $F^{2}$ kinetic
term multiplied with the $z$-axis thickness $d$. The final gauge
field action is \bea S_{em}\up=\int{\rm d}^3x
\left\{-\frac{d}{4}F^{\mu\nu}F_{\mu\nu}+\frac{\theta}{4}
\varepsilon_{\mu\nu\lambda}A^{\mu}F^{\nu\lambda}\right\},\eea

\noi where $\theta(x)=\sigma_{xy}$ in the bulk of the CDDW, which
is considered homogeneous. Variation of $S_{em}\up$ yields
\bea\partial^{\ph\nu}F_{\nu\mu}+\frac{\theta}{2d}\varepsilon_{\mu\nu\lambda}^{\phd}F^{\nu\lambda}
&=&-\frac{1}{2d}\varepsilon_{\mu\nu\lambda}^{\phd}(\partial^{\nu}\theta)A^{\lambda},\label{eq:motion}\\
n^i\left(\left.F_{i\lambda}^{\phd}\right|_{\sigma_+}-\left.F_{i\lambda}^{\phd}\right|_{\sigma_-}\right)&=&+\left.\frac{\theta}{2d}
A^{\mu}n^i\varepsilon_{\mu
i\lambda}\right|_{\sigma},\label{eq:boundary}\eea

\noi where $\bm{n}$ is the unit vector normal to the boundary
surface $\sigma$. Eq.(\ref{eq:motion}) describes the dynamics of
$A^{\mu}$ while Eq.(\ref{eq:boundary}) provides the boundary
conditions. In both equations the terms on the right hand side
stem from the fact that we are dealing with a bounded system and
the Chern-Simons action is not gauge invariant on the boundary
surfaces. In such cases, gauge invariance is recovered by current
carrying chiral edge modes \cite{Wen}. In the rest, we assume the
such modes do exist and extinguish the right hand sides of
Eq.(\ref{eq:motion}),(\ref{eq:boundary}).

Using the Coulomb gauge, $\bm{\nabla}\cdot \bm{A}=0$, and
considering the \textit{static} limit, we obtain the following
equations
\begin{figure}[t]\centering
\includegraphics[width=0.48\textwidth]{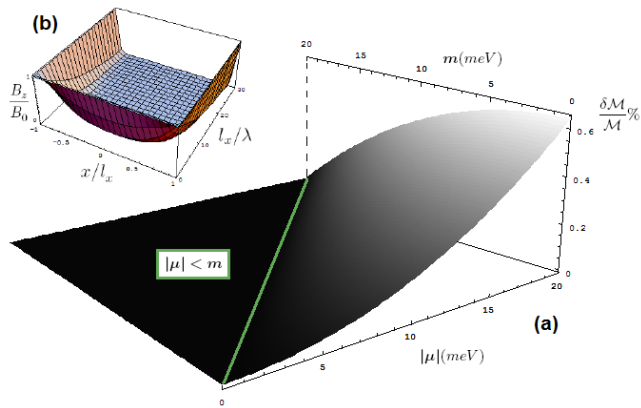}
\caption{(Color online) (a) Influence of doping on the Topological
Meissner effect. The relative change of magnetization hardly
reaches 1\% in the presence of a small chemical potential
($\Delta=20meV$, $t=500meV$). (b) The magnetic field screening as
a function of the position on the Hall bar extending from $-l_x$
to $+l_x$, for different values of the penetration depth $\lambda$
over $l_x$. The magnetic field is totally expelled from the sample
when $l_x/\lambda>>1$ exactly as in the superconducting
case.}\label{fig:TME}
\end{figure}
\bea\varepsilon\frac{\partial E_x}{\partial
x}-\frac{\sigma_{xy}}{d} B_z=0\quad and\quad
\frac{1}{\mu}\frac{\partial B_z}{\partial x}-\frac{\sigma_{xy}}{d}
E_x=0,\label{eq:EB}\eea

\noi where we have included the electric permittivity
$\varepsilon$ and the magnetic permeability $\mu$. To obtain the
TME, we apply a magnetic field of magnitude $B_0$ perpendicular to
the sample . The corresponding boundary conditions are $B_z(\pm
l_x)=B_0$. The magnetic field satisfies the differential equation
$\left(\frac{\partial^2}{\partial
x^2}-\frac{1}{\lambda^2}\right)B_z^{\phd}=0$, with
$\lambda=[(\sqrt{\varepsilon/\mu}\ph)/\sigma_{xy}]d$, the zero
temperature penetration depth. This is indeed the equation we find
in the case of a superconductor. Notice that in our case only the
$z$-component of the magnetic field is involved, implying that the
\textit{TME takes place only for magnetic fields perpendicular to
the plane}. Solving the above equation using the aforementioned
boundary conditions yields $
B_z(x)=B_0^{\phd}[\cosh(x/\lambda)]/{\cosh
(l_x/\lambda)}\label{eq:BB}$. \noi In Fig.(\ref{fig:TME}b) we plot
the magnetic field versus the ratio of $l_x/\lambda$ throughout
the whole sample. For $l_{x}>>\lambda$ we have almost complete
screening of the magnetic field. Integration of $B_{z}$
 over the whole sample yields the magnetization
${\cal M}={\cal
M}_{0}^{\phd}[\tanh\left({l_x^{\phd}}/{\lambda}\right)]/
\left({l_x^{\phd}}/{\lambda}\right)$, where we have introduced the
zero temperature magnetization ${\cal M}_0=B_0L_xL_y$,
corresponding to full penetration. One can readily obtain the
dependence of the magnetization on the chemical potential in the
$l_x\up>>\lambda$ regime. Using Eq.(\ref{eq:nex}), we find ${\cal
M(\mu)}\simeq{\cal M}_0\lambda(\mu)/l_x\up={\cal M}/(1-n_{ex})$
and $\delta{\cal M}/{\cal M}=n_{ex}/(1-n_{ex})$. As shown in
Fig.(\ref{fig:TME}a), the relative change of the magnetization due
to doping is unimportant for low doping, where it stands
$\delta{\cal M}/{\cal M}\simeq n_{ex}\sim \mu^2/t\Delta$.

\begin{figure}[t] \centering
\includegraphics[width=0.38\textwidth]{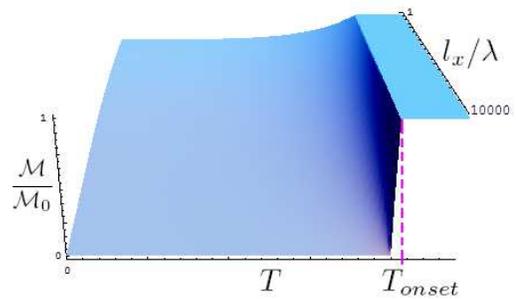}
\caption{(Color online) Magnetization ratio ${\cal M}/{\cal M}_0$
versus temperature $T$ for different values of the ratio
$l_x/\lambda$. For $T<T_{onset}$ the system is in a chiral
d-density wave state. In a BCS approximation, these curves are
identical to the superconducting case.} \label{fig:Magnetization}
\end{figure}

We may also obtain the temperature dependence of the magnetization
by assuming that the penetration depth has the usual BCS
temperature dependence
$[\lambda/\lambda(T)]^2={1-\left({T}/{T_{onset}}\right)^4}$. Under
this consideration, we obtain the magnetization curves shown in
Fig.(\ref{fig:Magnetization}) which are identical to the ones
encountered in the superconducting case as expected within this
BCS treatment. If we compare the magnetization curves of
Fig.(\ref{fig:Magnetization}) with the experimental results in the
cuprates, we observe that our BCS approximation does not provide a
fully satisfactory fit. Nonetheless, a strict quantitative
comparison calls for an implementation of our picture, that would
be more adapted to the cuprate materials.

\begin{figure}[!tl] \centering
\includegraphics[width=0.34\textwidth]{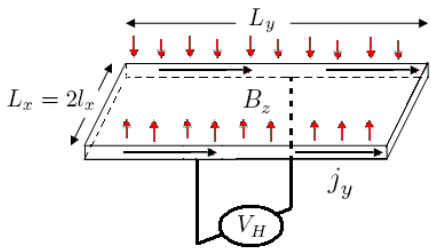}
\caption{ (Color online) Spontaneous Quantum Hall Effect as a
signature of the chiral d-density wave. Symmetrical surface
currents running along the y-axis create a magnetic field
distribution $B(\pm l_x)=\pm B_0$ leading to the spontaneous
generation of a Hall voltage $V_H$.}\label{fig:SQHE}
\end{figure}

A direct experimental verification of the presence of a CDDW could
be provided by the SQHE (Fig.(\ref{fig:SQHE})). If a CDDW is
present, a Hall voltage can be generated by the sole application
of a magnetic field. Specifically, we solve Eq.(\ref{eq:EB}) with
boundary conditions $E_{x}(\pm l_x)=0$ and $B(\pm l_x)=\pm B_0$,
as in Ref.\cite{Furusaki}. The spontaneously generated Hall
voltage is,
$V_{H}=vB_0L_x\left[\coth\left({l_x}/{\lambda}\right)-{\lambda}/{l_x}\right],$
where $v=1/\sqrt{\mu\varepsilon}$ is the velocity of light in the
material. If $\l_x>>\lambda$, $V_H^{\phd}=vB_0^{\phd}L_x^{\phd}$,
which implies that the applied magnetic field totally transforms
to an electric field. However, if there is a number of domains of
different chiralities in the sample, the SQHE is not an efficient
probe of the CDDW. Nevertheless, there is an alternative route in
detecting a CDDW. We refer to the gapless chiral edge modes that
exist on the boundary surfaces separating the bulk from the
vacuum. In order to restore the gauge invariance of the
Chern-Simons terms, these current carrying modes appear,
constituting a direct indication of a CDDW
\cite{Furusaki,{Horovitz}}.

Based on the ${\cal P-T}$ violation and the possibility of a
spontaneous electric Hall response via the SQHE we naturally
expect a CDDW to exhibit a spontaneous thermoelectric Hall
response \cite{Zhang}, detectable in principle, in a Nernst
measurement.
%
%In a Nernst setup, a thermal gradient along one direction and a
%perpendicular magnetic field are applied to the sample and the
%voltage in the transverse direction (Nernst signal) is measured.
%If a CDDW is present, there should be a contribution to the Nernst
%signal due to the simultaneous application of both external fields
%and another spontaneous contribution produced by the \textit{sole
%application of the thermal gradient}.
The unusual Nernst
contribution has the same origin with the TME and consequently
they should scale. This implies that the simultaneous presence of
both enhanced Nernst and diamagnetic signals reported in the
pseudogap regime is compatible with the assumption of a CDDW
state.

In conclusion, we have proposed an alternative way of generating a
Meissner effect without invoking in any manner superconductivity.
%This is particularly relevant for the pseudogap regime, in which,
%enhanced diamagnetic signals have been observed above the
%superconducting $T_c\up$.
In our picture, the existence of a chiral d-density wave (CDDW)
generates the Topological Meissner effect (TME) due to ${\cal
P-T}$ violation. The direction of the chirality of the CDDW
guarantees that the TME takes place only for perpendicular to the
plane magnetic fields, which is in agreement with the diamagnetic
observations in the pseudogap regime of the cuprates. Moreover, a
spontaneous thermoelectric response that accompanies the TME is
consistent with the observed unusual Nernst signal. Note also,
that the presence of a CDDW is compatible with the recently
observed quantum oscillations in YBCO \cite{quantum oscillations}
that reported Fermi surface pockets in the nodal areas, possibly
indicating the doubling of the Brillouin zone. As a matter of
fact, associating a CDDW with the pseudogap regime seems quite
promising and undoubtedly further theoretical investigation should
be performed.
%As far as the experimental probes of a CDDW are
%concerned, the SQHE and the gapless chiral edge modes constitute
%definite experimental signatures, provided they become observable.

We are grateful to Professors P. B. Littlewood, N. P. Ong, V.M.
Yakovenko, M. Sigrist and P. Thalmeier for stimulating and
enlightening comments and discussions. We acknowledge financial
support by the EU through the STRP NMP4-CT-2005-517039 CoMePhS.
One of the authors (P.K.) also acknowledges financial support by
the Greek Scholarships State Foundation.

\end{document}